# Experiments on Crowdsourcing Policy Assessment

JOHN PRPIĆ            - Beedie School of Business, Simon Fraser University
ARAZ TAEIHAGH[1]      - City Futures Research Centre, University of New South Wales
JAMES MELTON          - College of Business Administration, Central Michigan University

--------------------------------------------------------------------------------------------------

**Abstract**

Can Crowds serve as useful allies in policy design? How do non-expert Crowds perform relative to experts in the assessment of policy measures? Does the geographic location of non-expert Crowds, with relevance to the policy context, alter the performance of non-experts Crowds in the assessment of policy measures? In this work, we investigate these questions by undertaking experiments designed to replicate expert policy assessments with non-expert Crowds recruited from Virtual Labor Markets. We use a set of ninety-six climate change adaptation policy measures previously evaluated by experts in the Netherlands as our control condition to conduct experiments using two discrete sets of non-expert Crowds recruited from Virtual Labor Markets. We vary the composition of our non-expert Crowds along two conditions: participants recruited from a geographical location directly relevant to the policy context and participants recruited at-large. We discuss our research methods in detail and provide the findings of our experiments.

## 1. Introduction

Crowds of non-experts accessed through a form of Crowdsourcing (de Vreede et al 2009, Prpić, Taeihagh, and Melton, 2014) known as Virtual Labor Markets (VLMS) have been shown to perform exceptionally-well relative to experts in several areas. These areas include, for example, image segmentation (Lee 2013), photograph classification (Mitry et al 2013), and biomedical ontologies (Mortenson et al 2013). Further, a growing body of experimental work (Micallef et al 2012, Trushkowsky et al 2013, Bragg and Weld 2013, Lasecki et al 2013, Hossfeld et al 2013) undertaken through VLMs such as Amazon's M-Turk,

---

[1] Corresponding author: araz.taeihagh@new.oxon.org, City Futures Research Centre, FBE, UNSW, Sydney 2052, NSW, Australia.

Crowdflower, and Gigwalk, illustrates that experiments with participants at VLMs are equal to or better than laboratory experimentation in respect to participant diversity and overall rigour (Paolacci et al 2010, Yen et al 2013).

Given these early precedents illustrating that non-experts can perform well relative to experts in a number of domains and that experiments at VLMs have been shown to be scientifically viable settings for rigorous experimentation, in this work we design, implement, and report results from two VLM-based replication experiments for policy assessment.

In the ensuing sections of this work, we first briefly review the literature on Policy Design and Crowdsourcing to motivate the context of our research and contributions. From this platform, we then describe our experimental design and data collection procedures before presenting the results of our findings.

## 2. Policy Design

Due to inherent technical, institutional, and political difficulties, policy problems are often considered to be 'messy' (Ney, 2009) or 'wicked' (Rittel and Webber, 1973). No longer are policy makers faced with a lack of options when dealing with policy problems; rather, sufficiently exploring all of the viable options has become more difficult and time consuming. In the 21st century, the number of policy alternatives and their respective measures considered in a policy problem can number in the hundreds, and as a result what should be done and what should be done first is becoming an increasingly complex question, given the options available, the information (empirical and/or theoretical) on each option, and the various political and advocacy influences on policy making (Taeihagh et al. 2013). Furthermore, there is evidence for inertia and a lack of consideration of more than a few options (Kelly et al., 2008).

To successfully and efficiently advance most complex policy problems, using a range of different measures (i.e. a policy package - see Givoni et al, 2010) is necessary. One of the challenging tasks in executing such a methodology is processing a vast amount of information and developing a library of measures with the accompanying assessment of their properties. To facilitate policy design, new methodologies and decision support systems have been proposed (Taeihagh et al., 2014), and in this vein, this work seeks to investigate the performance of non-expert Crowds from VLMs in relation to the performance of known-experts in the evaluation of policy alternatives. In broad-terms, we seek to evaluate whether a Crowd can be a useful ally in policy design. Our results may have important implications in policy design in general, and specifically in the evaluation of policy packages, as the use of non-expert Crowds from VLMs has the potential to reduce the time and cost of analysis and may support the validation of policy measures.

# 3. Crowdsourcing

Crowdsourcing is known to be a distributed problem-solving, idea generation, and production model where problems are broadcast through IT to an unknown group in the form of an open call for input (Brabham 2008). Problem-solving, idea-generation and production are generally achieved by different IT-mediated approaches, and in the following subsections we will describe the three general types of Crowdsourcing found in the literature (de Vreede et al 2013).

## 3.1 Tournament-Based Crowdsourcing

In tournament-based crowdsourcing (TBC), organizations post their problems or opportunities to IT-mediated Crowds at web properties such as Innocentive, Eyeka, and Kaggle (Afuah and Tucci 2012). In posting a problem at the web property (and paying for the privilege to do so), the organization creates a prize competition amongst the assembled Crowd where the best solution will be chosen (and granted the prize) as determined by the sponsor organization.

These web properties leverage an exclusive Crowd of contributors that they have formed and offer client organizations access to their Crowd in exchange for fees (Zwass 2010, Lakhani & Panetta 2007). These web properties generally attract and maintain more or less specialized Crowds premised upon the focus of the web property, though the properties are almost always open to new Crowd members joining. For example, the Crowd at Eyeka is coalesced around the creation of advertising collateral for brands, while the Crowd at Kaggle has formed around data science (Ben Taieb and Hyndman 2013, Roth and Kimani 2013). When applied to innovation, these platforms have been termed as innomediaries or open innovation platforms (Sawhney et al. 2003) and represent both the idea generation and problem solving aspects of Crowdsourcing (Morgan and Wang 2010, Brabham 2008).

The scale of the Crowd of participants at these TBC sites can be very large (for example, Kaggle has a Crowd of approximately 140,000 available, while eYeka boasts approximately 280,000 members), and the individual Crowd participants can choose whether or not to be anonymous at these sites in relation to their offline identities.

Generally, firms seek out such TBC sites for relatively complete solutions (for example a new algorithm that solves a business need) and the competition sponsoring firm usually works directly with the TBC intermediary to design the contest in respect its duration, determine the number of prizes being offered, the monetary distribution of the prize money amongst the overall prizes awarded, and define the problem to be solved itself. Fixed amounts of prize money are offered to the Crowd for the winning solution and can range from a few hundred dollars to a million dollars or more[2].

## 3.2 Open Collaboration

In the open collaboration (OC) model of Crowdsourcing, organizations post their problems/opportunities to the public at large through IT. Contributions from the Crowds in these endeavors are voluntary and do not generally entail monetary exchange. Posting on Reddit, starting an enterprise wiki (Jackson and Klobas 2013), or using social media (Kietzmann et al 2011) like Facebook and Twitter (Sutton et al 2014) to garner contributions are prime examples of this type of Crowdsourcing. In this vein, Wikipedia is perhaps the most famous example of an OC application, "where a huge amount of individual contributions build solid and structured sources of data" (Prieur et al. 2008).

The scale of the Crowds available to these types of endeavors can vary significantly depending on the reach and engagement of the IT used and the efficacy of the "open call" for volunteers. For example, Twitter[3] currently counts 271 million active users per month, with 500 million tweets sent on an average day, while applications like Reddit have approximately 3 million registered 'Redditors'.[4] Further, individuals such as politicians like Narendra Modi of India have in place very large personal communities of followers on Facebook to the tune of 15 million 'likes'[5]. On the other hand, California Assemblyman Mike Gatto garnered less than one hundred responses in his efforts to Crowdsource probate legislation in California.[6]

In short, the size of the Crowds accessed in OCs can vary significantly in terms of the upper limit of Crowd scale, easily varying orders of magnitude at a time, depending on the particular platform in use. Further, even with the largest potential Crowds at, say, Facebook or Twitter, there is little to guarantee the attention of any significant subset of the vast amount of potential contributors at any one time. Similarly, in respect to anonymity, OC Crowds run the gamut from fully anonymous--in respect to one's offline identity--to participation that is completely synonymous with offline identity.

## 3.3 Virtual Labor Marketplaces

A virtual labor marketplace (VLM) is an IT-mediated market for spot labor where individuals and organizations can agree to execute work in exchange for monetary compensation. Typified by endeavors like Amazon's M-Turk and Crowdflower, these endeavours are generally thought to exemplify the "production model" aspect of Crowdsourcing (Brabham 2008), where workers undertake microtasks for pay. Microtasks, such as the translation of documents, labelling/tagging photos, and transcribing audio (Narula et al. 2011), are generally considered to represent forms of human computation (Michelucci 2013, Iperiotis & Paritosh 2011), where human intelligence is tasked to undertake work currently unachievable through artificial intelligence means. Though human computation tasks cannot be tackled by the most advanced forms of AI, they are generally rather mundane work for all human intelligences.

---

[2] http://www.innocentive.com/files/node/casestudy/case-study-prize4life.pdf.
[3] https://about.twitter.com/company
[4] http://www.reddit.com/about
[5] https://www.facebook.com/narendramodi
[6] http://mikegatto.wikispaces.com

In respect to the Crowd of laborers at VLMs, they are generally anonymous in respect to their offline identities (Lease et al 2013) and independently undertake the tasks that they select, based upon the compensation offered for the task and the nature of the task itself. Crowd laborers accrue historical ratings for their past efforts with tasks, which future requesters can use to segment the highest quality workers. In this respect, the work is completed by Crowd participants first, and then the requesters (or the platform intermediary) decide whether or not to pay for the work, based upon their satisfaction with the effort. The size of the Crowds at the VLMs can be immense; for example, Crowdflower has over 5 million laborers available at any given time. Therefore, microtasking through VLMs can be rapidly completed on a massively parallel scale.

With our review of the Crowdsourcing literature complete, in the next section we describe our research methods in using VLM Crowdsourcing to garner policy assessments from separate Crowds of non-experts.

## 4. Research Method

In this section we outline the details of our research method, including our overall experimental design, our experimental task design, the pilot tests undertaken, the final experiments, and the data collected.

### 4.1 Experimental Design

Using a deterministic climate change scenario (see Appendix #1) and an accompanying set of ninety-six policy measures (see Appendix #2 for a sample) for climate change adaptation previously evaluated by experts in the Netherlands (de Bruin et al 2009), we present the identical climate change scenario and adaptation policy measures--as our control condition-- to two discrete sets of non-expert Crowds recruited from Virtual Labor Markets (VLMs).

We vary the composition of our non-expert Crowds along two conditions: participants recruited from a geographical location directly relevant to the policy context (i.e., the Netherlands) and participants recruited at-large. In each experiment, we gather approximately 25 independent assessments for each of the ninety-six climate change adaptation options along five criteria (importance, urgency, no-regret, co-benefit, mitigation) on a 5 point scale.

**4.1.1 Experimental Task Design**

Our experimental tasks give the participants a climate change scenario in a deterministic setting, in which it is assumed that the climate changes detailed in Appendix #1 will occur by the year 2050 in the Netherlands. Within this deterministic setting, participants are asked to assess different policy measures to prepare for these assumed climate changes. Policy measures are evaluated along a solitary set of assessments for adaptation priority (see Appendix #3). The adaptation priority assessment set contains five independent criteria (importance, urgency, no-regret, co-benefit, mitigation), each assessed on a 5-point scale ranging from very low (1) to very high priority (5).

The ninety-six policy measures that were evaluated by experts in de Bruin et al (2009) were evaluated as a complete corpus by each expert, whereas in our experiment we break the body of ninety-six policy measures into ninety-six separate microtasks, each containing exactly one policy measure. Appendix #4 illustrates screenshots of an actual task implemented during the experiments.

After receiving IRB approval but prior to executing the main experiments, we undertook a set of iterative pilot tests at the M-Turk and Crowdflower VLM platforms to refine the text and presentation of our experimental tasks and to find adequate pricing-levels for our tasks in the markets. A total of eight different pilot tests were initiated and iterated altogether.

Initially, we thought that we would be able to run both of our experiments (Dutch and at-large) at the M-Turk platform alone, though it soon came to light from our pilot tests that it would not be possible to find and recruit a solely Dutch Crowd of participants at M-Turk, irrespective of the reward offered. Therefore, we turned to Crowdflower to run our Dutch experiment.

The net result of our complete set of pilot studies was attractively displayed microtasks at each VLM platform, appropriately structured with task-settings such as price, expiry of task, and maximum time spent per task, etc. Further, the pilots gave us a working knowledge of the form of the output of the results of Crowd labor and some insight into the market dynamics of VLM Crowd behaviour.

Altogether, the ability to rapidly iterate through pilot tests at VLMs is, we feel, a powerful strength of the methodology. After running the pilot tests for seven days, we were able to fine tune the experiments and run the main experiments tasks given the gained understanding of the time, monetary costs and market dynamics of the VLMS for the proposed microtasks.

### 4.2 Main Experiments

Over a period of approximately 3 weeks bridging July and August 2014, our final experiments were undertaken. As mentioned, the Dutch experiment was run at the Crowdflower platform, while the at-large experiment was run at M-Turk. At both Crowdflower and M-Turk, ninety-six different "batches" were created and launched through the web interface, each reflecting one climate change adaptation option.

Each task was typically completed with the lower bound of 15 seconds to the upper bound of 2 minutes for majority of the participants. With average times ranging from 30 to 45 seconds per task. Compensation of one to five cents per task were offered at Crowdflower and M-Turk after conducting pilot tests to understand market dynamics for the microtasks defined. The Dutch experiment was limited to participation of those who were verified by Crowdflower to be Dutch residents. Participation in the M-Turk batches was not restricted geographically

### 4.3 Data Collected

Our Dutch experiment yielded twenty five completed assessments (25) for each of the five criteria (5) for each climate change adaptation option (96). Our at-large experiment yielded an average of twenty-three-and-a-half completed assessments (23.5), for each of the five criteria (5), for each climate change adaptation option (96). Exactly half of the at-large experiment climate change adaptation options received a full-set of twenty five completed assessments, while the remainder received between 16-24 assessments per climate change adaptation option. Altogether, 12,000 data points were collected from the Dutch experiment (25 X 5 X 96), while 11, 280 data points were collected from the at-large experiment (23.5 X 5 X 96). The cost to collect these 23,000+ data points was approximately $250 USD.

Crowd diversity has been shown by Hong & Page (2002) to be a fundamental element in the efficacy of IT-mediated Crowds. Because the microtasks were structured to contain only one climate change adaptation option at a time (as opposed to each expert assessing all 96 of them at once), our collected data represents a significant increase in overall respondent diversity, as compared to the expert assessments. Through our methods, it was possible for 25 discrete individuals at each VLM to evaluate each of the 96 adaptation options, therefore equaling a maximum participant pool of 2400 individuals supplying only one solitary assessment each. Though these details are beyond the scope of this particular work, we estimate that approximately 10% of Crowd members supplied a full set of 96 assessments in both experiments, which means that the overall participant pool (in supplying at least one assessment) for each experiment, contained at least one input from more than 230 different people (on average), representing input from people more than an order of magnitude higher than the expert assessments.

## 5. Results

Our research design was formed as an attempt at two replication experiments for policy assessment. We used the deterministic climate change scenario and the ninety-six related climate change adaptation options used by de Bruin et al (2009) to gather expert assessments on climate change adaptation and verbatim brought the exact same scenario and climate change adaptation options to two different Crowds of non-experts for assessment.

The use of the exact same scenario and adaptation options, scale, and criteria of evaluation used in the work of de bruin et al (2009) with our two different groups of non-expert Crowds, suffices as the control condition of our experiments, hence potentially allowing the replication of the expert policy assessment results, in both new non-expert Crowd settings.

Our experimental treatment varies the composition of the non-expert Crowds that we access to attempt to tease out if the composition of non-expert Crowds (as reflected by the geographic diversity of Crowd participants) changes the policy assessment results. To control for this treatment, one non-expert Crowd was recruited at-large, while a second distinct non-expert Crowd was recruited from a specific geographic setting relevant directly to the policy context (i.e., the Netherlands).

### 5.1 Results from Non-Expert Dutch Crowd

Our Dutch experiment with non-experts yielded twenty five completed assessments (25), for each of the five criteria (5), for each climate change adaptation option (96), forming 12,000 data points from this sample. And based upon our estimate that 10% of respondents completed all 96 assessments (and thus on average each respondent completed approximately 10 assessments each), our data points represent input from approximately 240 Dutch people overall.

As mentioned, this Dutch Crowd was accessed at the Crowdflower VLM platform, and we were able to segment this Crowd specifically and completely as a part of the functionality that the Crowdflower platform provides. Respondents, segmented based upon their IP addresses, were shown to originate from every region of the country, including urban and rural areas.

In respect to the actual assessments made by the Dutch Crowd of non-experts, Table #1

details a rank ordering of the Top 15 Assessments made by this Dutch Crowd using the Median values from the data collected and the same weights assigned by de Bruin et al. (2009).

**Table 1 - Top 15 Adaptation Options as Generated By Dutch Non-Expert Crowd**

| Category | Adaptation Option |
|---|---|
| Agriculture | Adjusting crop rotation schemes and planting and harvesting dates. |
| Agriculture | Self-sufficiency in production of roughage. |
| Agriculture | Water management and agriculture. |
| Agriculture | Choice of crop variety and genotype. |
| Agriculture | Development and growing of crops for biomass production. |
| Agriculture | Irrigation. |
| Agriculture | Floating greenhouses. |
| Agriculture | Limiting the import of timber. |
| Agriculture | Adjusting fishing quota. |
| Agriculture | Introduction of ecosystem management in fishery. |
| Nature | Afforestation and mix of tree species. |
| Energy & Transport | Use improved opportunities for generating wind energy. |
| Agriculture | Water storage on farmland. |
| Agriculture | Changes in farming systems. |
| Agriculture | Adaptation strategies to salinization of agricultural land. |

## 5.2 Results from Non-Expert At-Large Crowd

Our At-large experiment with non-experts yielded an average of twenty-three and a half completed assessments (23.5) for each of the five criteria (5) for each climate change adaptation option (96), forming 11,280 data points from this sample. Exactly half of the at-large experiment climate change adaptation options received a full-set of twenty five completed assessments, while the remainder received between 16-24 assessments per climate change adaptation option. And based upon our estimate that 10% of respondents completed all 96 assessments (and thus on average each respondent completed approximately 10 assessments each), our data points represent input from approximately 230 at-large people overall.

In respect to the actual assessments made by the at-large Crowd of non-experts, Table #2 details a rank-ordering of the Top 15 Assessments made by this Crowd. In identical fashion to the directly preceding analysis, our statistical aggregation of the values from the assessment data used the median values from the data collected and then sequenced the Top 15 median values to create the rank-ordered list seen in Table #2.

**Table #2 - Top 15 Adaptation Options as Generated by At-Large Non-Expert Crowd**

| Category | Adaptation Option |
|---|---|
| Energy & Transport | Use improved opportunities for transport generating solar energy. |
| Agriculture | Water management and agriculture. |
| Water | Protection of vital infrastructure. |
| Energy & Transport | Change modes of transport and develop more intelligent infrastructure. |
| Nature | Implementation of effective agri-environmental schemes. |
| Nature | Educational programs. |
| Energy & Transport | Water management systems: revision of sewer system. |
| Water | More space for water: Regional water system improving river capacity. |
| Water | Creating public awareness. |
| Energy & Transport | Use improved opportunities for generating wind energy. |
| Energy & Transport | Design houses with good climate conditions (control)—'low energy'. |
| Housing & Infrastructure | Water management systems: options for water storage and retention in or near city areas. |
| Housing & Infrastructure | Water management systems: emergency systems revision for tunnels and subways. |
| Nature | Integrated coastal zone management. |
| Water | Protection of vital objects. |

## 7. Conclusion

This work examines the use of crowdsourcing in policy design specifically by posing questions in regards to viability of non-expert crowds recruited through VLMs in aiding in policy design process. In our work we briefly reviewed policy design as one of the steps in the policy cycle and introduced and examined crowdsourcing approaches, focusing on VLMs and their unique characteristics. Using a previously conducted expert assessment of climate change adaptation policy measures as a benchmark, we created pilot and final experiments with two sets of crowds; one local to the specific policy context and an at large crowd, and we present the details of our experimental process. Thereafter, we present the results from our experiments.

## Acknowledgements
We would like to provide our sincere thanks and grateful acknowledgement to Prof. van Ierland and de Bruin et al. (2009) research team for allowing us to use their Climate Change Adaptation options and expert policy assessment results which allowed this work to be systematically undertaken as reported here.

# Appendix #1 - Deterministic Climate Change Scenario in The Netherlands by 2050[7]

|  | Estimated Change |
|---|---|
| Temperature (∘C) | +2 |
| Average summer precipitation (%) | +2 |
| Average summer evaporation (%) | +8 |
| Average winter precipitation (%) | +12 |
| Sea level rise (cm) | +60 |

# Appendix #2 - Assessment Set & Criteria of Climate Change Adaptation Option

| Assessment Set | Criteria |
|---|---|
| Adaptation | - Importance |
|  | - Urgency |
|  | - No Regret |
|  | - Co-Benefit |
|  | - Mitigation Effect |

# Appendix #3 - Sample of Policy Measures Used

a) Change modes of transport and develop more intelligent infrastructure
b) Improvement of healthcare for climate related diseases
c) Make existing and new cities robust—avoid 'heat islands', provide for sufficient cooling capacity
d) Artificial reefs along the coastline & development nature conservation values
e) Relocation of fresh water intake points
f) Adapted forms of building and construction
g) Increasing genetic and species diversity in forests
h) Eco-labelling and certification of fish
i) Adjusting crop rotation schemes and planting and harvesting dates
j) Floating greenhouses

---

[7] Adapted from de Bruin et al (2009) and KNMI (2003).

# Appendix # 4 - Sample VLM Task Screenshots - M-Turk

**Instructions**

### Scenario

By the year 2050, the Netherlands is expected to experience a change in climate.
The values below represent the most likely scenario for the coming changes, relative to today's climate measurements.

| | Estimated Change |
|---|---|
| Average temperature (°C) | +2 degrees Celsius |
| Average summer precipitation (%) | +2 % |
| Average summer evaporation (%) | +8 % |
| Average winter precipitation (%) | +12% |
| Sea level rise (cm) | +60 centimetres |

Please note that:
- 2 degrees Celsius = 3.6 Fahrenheit
- 60 centimetres = 23.6 inches

For the climate change scenario described above, please provide your best judgment below (without the use of any other materials, including the Internet/web etc) assessing the climate change adaptation option listed below, along the criteria presented.
A climate change adaptation option is an action that can be taken by government to adapt to the climate changes explained in the above scenario.

### Climate Change Adaptation Option

Water management systems: for water storage and retention in or near city areas.

1. **Importance** - in terms of the expected gross benefits that can be obtained from the option.
Select one of the following.

[ 1 (Low Priority) ▼ ]

2. **Urgency** - the need to act sooner rather than later on the option.
Select one of the following.

[ 1 (Low Priority) ▼ ]

3. **No-Regret** - is the option useful to implement irrespective of climate change?
Select one of the following.

[ 1 (Low Complexity) ▼ ]

4. **Co-Benefit** – does this option benefit other sectors and domains?
Select one of the following.

[ 1 (Low Priority) ▼ ]

5. **Mitigation Effect** – does this option mitigate climate change?
Select one of the following.

[ 1 (Low Priority) ▼ ]